\documentclass[
reprint,
amsmath,amssymb,
aps,
showkeys
]{revtex4-2}
\usepackage{graphicx}
\usepackage{dcolumn}
\usepackage{bm}
\usepackage{hyperref}
\hypersetup{colorlinks,allcolors=violet}
\usepackage{float}
\usepackage[braket, qm]{qcircuit}
\usepackage{multirow}
\usepackage{booktabs} 
\usepackage{siunitx} 
\usepackage{xcolor}
\usepackage[table]{xcolor}
\usepackage{amsmath}

\begin{document}

\title{Pattern-Dependent Performance of the Bernstein-Vazirani Algorithm}

\author{Muhammad AbuGhanem{$^{1,2}$}}
 \address{$^{1}$ Faculty of Science, Ain Shams University, Cairo, $11566$, Egypt}
\address{$^{2}$ Zewail City of Science, Technology and Innovation, Giza, $12678$, Egypt}

\email{gaa1nem@gmail.com}

\date{\today}

\begin{abstract}

Quantum computers promise to redefine the boundaries of computational science, offering the
potential for exponential speedups in solving complex problems across chemistry, optimization, and
materials science. Yet, their practical utility remains constrained by unpredictable performance degradation under real-world noise conditions. A key question is how problem structure itself influences algorithmic resilience. In this work, we present a comprehensive, hardware-aware benchmarking study of the Bernstein-Vazirani algorithm across 11 diverse test patterns on multiple superconducting quantum processors, revealing that algorithmic performance is exquisitely sensitive to problem structure. Our results reveal average success rates of 100.0\% (ideal simulation), 85.2\% (noisy emulation), and 26.4\% (real hardware), representing a dramatic 58.8\% average performance gap between noisy emulation and real hardware execution. With quantum state tomography confirming corresponding average state fidelities of 0.993, 0.760, and a 0.234 fidelity drop to hardware. Performance degrades dramatically from 75.7\% success for sparse patterns to complete failure for high-density 10-qubit patterns. Most strikingly, quantum state tomography reveals a near-perfect correlation (r = 0.972) between pattern density and state fidelity degradation, providing the fundamental explanation for observed performance patterns. The catastrophic fidelity collapse observed in real hardware measurements—dropping to 0.111 compared to the predicted 0.763—underscores severe limitations in current noise models for capturing structure-dependent error mechanisms. These findings suggest that problem formulations should minimize entanglement density and avoid symmetric encodings to achieve viable performance on current quantum hardware. Our work establishes pattern-dependent performance as a critical consideration for quantum algorithm deployment and provides a quantitative framework for predicting algorithm feasibility in practical applications.

\end{abstract}


\maketitle

\tableofcontents

\section{Introduction}

Quantum computing represents a fundamental shift in computational paradigms~\citep{Light,SuperconductingQuantum}, leveraging quantum mechanical phenomena to solve problems that remain intractable for classical computers~\cite{Kim23,NISQ24,Willo25,IBMQuantum,PhotonicQuantumComputers}. By exploiting superposition, entanglement, and interference, quantum algorithms offer exponential speedups for specific problem classes, from factoring large integers~\citep{Shor,factor2048bitRSA} to simulating quantum systems~\citep{Cerezo2021}. This computational advantage~\citep{NISQ24} has profound implications across scientific domains, including drug discovery~\citep{Drugdesign}, materials science~\citep{Quantumchemistrysimulation}, neuroscience~\citep{Wolff2024quantumneuro} and optimization~\citep{supplychains}. As quantum hardware evolves through the Noisy Intermediate-Scale Quantum (NISQ) era~\citep{NISQ18} toward fault tolerance~\citep{Shor-Fault-tolerantQC}, understanding its practical capabilities and limitations is critical for realizing this potential.

Accurate noise models are essential for bridging the gap between theoretical promise and practical utility, serving as critical ``digital twins'' of quantum hardware for performance prediction and error mitigation~\cite{Kamakari2022, Cai2023,GoogleAI}. While significant progress has been made in modeling generic gate and readout errors, the relationship between algorithmic performance and specific problem structure remains poorly characterized. In particular, the impact of problem-specific structural properties—such as entanglement patterns and symmetry—remains largely unquantified, limiting our ability to predict which quantum algorithms will maintain their advantage under realistic noise conditions.

The Bernstein-Vazirani (BV) algorithm provides an ideal platform to investigate this challenge. Its journey from theory to practice is marked by critical developments. Early experimental implementations confirmed its quantum advantage in controlled trapped-ion systems~\citep{fallek2016implementation}, while theoretical work unlocked its potential for cryptanalysis~\citep{xie2019cryptanalysis,xu2024differential}. Resource-theoretic analyses have clarified that coherence, not entanglement, is the key driver of its speedup~\citep{naseri2022coherence}, a finding refined by the recent identification of the 'coherence fraction' as the precise quantitative resource~\citep{zhou2025coherencefraction}. However, the path to practical application is challenging, with studies demonstrating the algorithm's sharp performance decline under noise~\citep{gupta2023noise} and the exponential suppression of error rates required for scaling~\citep{faizan2025complexity}.

Yet, a critical gap persists. Existing noise models largely treat errors as independent of the specific computational problem being solved. The fundamental question of how the inherent structure of a problem—the very bit string pattern encoded into the BV oracle—dictates its resilience to real-world hardware noise has been largely unexplored.

In this work, we address this gap directly through a comprehensive benchmarking study on multiple 127-qubit superconducting quantum processors~\citep{IBMQuantum}. We demonstrate that the performance of the Bernstein-Vazirani algorithm is not merely a function of scale and generic error rates, but is exquisitely sensitive to problem structure. Our investigation reveals a pronounced performance hierarchy tied directly to pattern complexity, with success probabilities varying dramatically based on structural properties.

This work contributes a systematic benchmarking framework for evaluating quantum algorithms across diverse structural pattern classes, enabling performance analysis that moves beyond simple metrics like qubit count. Our findings further indicate that pattern density and symmetry are significant factors in performance degradation on NISQ hardware, suggesting important limitations in current noise models that overlook these structural dependencies. Finally, the study yields practical insights for structure-aware algorithm design, demonstrating through rigorous experimentation that a problem's specific encoding critically impacts its successful execution on real quantum systems. Together, these contributions offer a methodological foundation for more accurately assessing algorithm viability on contemporary noisy quantum devices.

By connecting structural features to performance outcomes, this work provides a crucial roadmap for navigating the NISQ era~\citep{NISQ18}, prioritizing problem formulations that are not only mathematically elegant but also structurally robust against the realities of noisy quantum hardware.

This paper proceeds as follows. 
Section~\ref{sec:Methodology} details our methodology, including the Bernstein-Vazirani algorithm implementation, experimental design, hardware platforms, and performance metrics. 
Section~\ref{sec:pattern-dependent-performance} presents our comprehensive results analyzing pattern-dependent performance across complexity classes. 
Section~\ref{sec:discussions} discusses the implications of our findings, particularly the entanglement-induced performance cliff. 
Section~\ref{sec:QST} validates these results through quantum state tomography, revealing the fundamental mechanisms behind performance degradation. 
Section~\ref{sec:implications} derives implications for algorithms design. Finally, Section~\ref{sec:conclusion} concludes with key insights and future research directions.

\section{Methodology}
\label{sec:Methodology}

This section outlines the experimental framework designed to systematically investigate the impact of problem structure on quantum algorithm performance. Our methodology encompasses the selection of a benchmark algorithm, the design of a diverse test suite, the implementations in diverse settings---quantum simulation, noisy emulation, and physical quantum hardware, and the use of quantum state tomography for validation.

\begin{figure*}
    \centering
    \includegraphics[width=\textwidth]{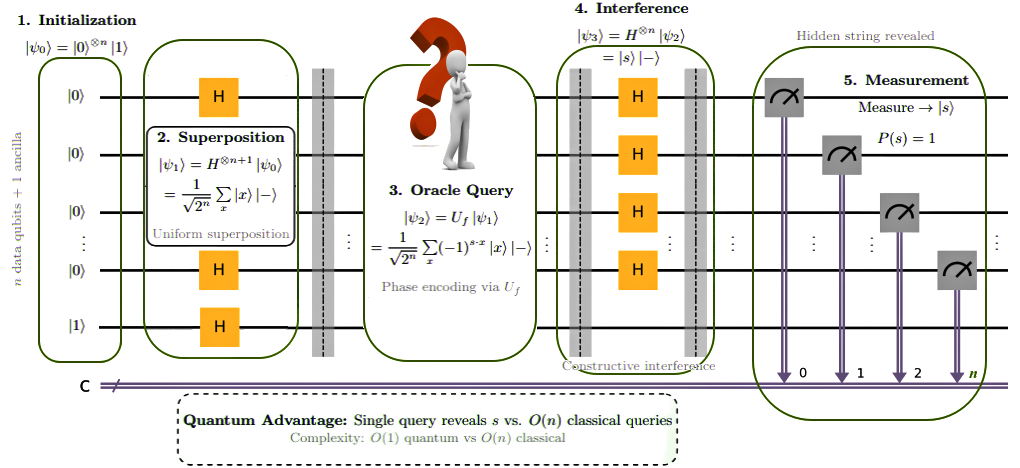}
\caption{Bernstein-Vazirani Algorithm. Step-by-step quantum state evolution demonstrating exponential quantum advantage. The algorithm progresses through five key transformations: 
(1) Initialization prepares $\ket{\psi_0} = \ket{0}^{\otimes n}\ket{1}$, 
(2) Superposition creates $\ket{\psi_1} = \frac{1}{\sqrt{2^n}}\sum_{x\in\{0,1\}^n} \ket{x}\ket{-}$ via Hadamard gates, 
(3) Oracle application encodes the hidden string through phase kickback $\ket{\psi_2} = \frac{1}{\sqrt{2^n}}\sum_x (-1)^{s\cdot x}\ket{x}\ket{-}$, 
(4) Interference via final Hadamard gates collapses the state to $\ket{\psi_3} = \ket{s}\ket{-}$, and 
(5) Measurement reveals $s$ with unit probability. The complete state evolution is mathematically characterized by:
\(
\ket{\psi_0} = \ket{0}^{\otimes n}\ket{1}, \quad 
\ket{\psi_1} = \frac{1}{\sqrt{2^n}}\sum_x \ket{x}\ket{-}, \quad 
\ket{\psi_2} = \frac{1}{\sqrt{2^n}}\sum_x (-1)^{s\cdot x}\ket{x}\ket{-}, \quad 
\ket{\psi_3} = \ket{s}\ket{-}
\)
This quantum workflow achieves $O(1)$ query complexity compared to $O(n)$ classically, providing exponential speedup for hidden string recovery.}
\label{fig:bv_circuit_cover}
\end{figure*}

\subsection{Benchmark Algorithm: Bernstein-Vazirani}
\subsubsection{Algorithm Steps and Implementation}

The Bernstein-Vazirani algorithm \cite{bernstein1993,bernstein1997} was selected as the benchmark for this study due to its conceptual simplicity, deterministic nature, and sensitivity to noise. The BVA solves the problem of identifying a hidden bitstring \( s = s_0s_1\ldots s_{n-1} \) encoded within a black-box oracle function \( f_s(x) = s \cdot x \mod 2 \). The function returns the dot product \( s \cdot x \) modulo 2 for an input \( x \). While a classical computer requires \( O(n) \) queries, the quantum algorithm finds \( s \) with a single query, providing a clear quantum advantage~\citep{bernstein1997,GSA}.

The quantum algorithm, consists of the following steps:
   
\begin{enumerate}
    \item \textbf{Initialization:} The system is initialized with \( n \) qubits are in the \( \ket{0}^{\otimes n} \) state and one ancillary qubit in the $\ket{1}$ state.
    \begin{equation}
    \ket{\psi_0} = \ket{0}^{\otimes n} \otimes \ket{1}
    \end{equation}
    
    \item \textbf{Superposition:} Hadamard gates are applied to all qubits, creating a uniform superposition across all input states, with the ancilla qubit in the $\ket{-}$ state.
    \begin{equation}
    \ket{\psi_1} = (H^{\otimes n} \otimes H) \ket{\psi_0} = \frac{1}{\sqrt{2^n}} \sum_{x \in \{0,1\}^n} \ket{x} \otimes \frac{\ket{0} - \ket{1}}{\sqrt{2}}
    \end{equation}
 
    \item \textbf{Oracle Application:} The oracle \( U_f \) is applied, which implements the function \( f_s(x) = s \cdot x \mod 2 \) by conditionally applying phase flips to computational basis states. For each position where \( s_i = 1 \), the oracle modifies the quantum state's phase based on the corresponding qubit value in \( |x\rangle \), creating interference patterns that encode information about \( s \).

     \begin{equation}
    \ket{\psi_2} = U_f \ket{\psi_1} = \frac{1}{\sqrt{2^n}} \sum_{x \in \{0,1\}^n} (-1)^{f_s(x)} \ket{x} \otimes \frac{\ket{0} - \ket{1}}{\sqrt{2}}
    \end{equation}
    The phase oracle encodes the function $f_s(x) = s \cdot x$ into the quantum state's phase:
    \begin{equation}
    U_f: \ket{x}\ket{y} \rightarrow \ket{x}\ket{y \oplus f_s(x)}
    \end{equation}
    For the $\ket{-}$ ancilla state, this results in phase kickback:
    \begin{equation}
    \ket{x}\ket{-} \rightarrow (-1)^{f_s(x)} \ket{x}\ket{-}
    \end{equation}
    
    \item \textbf{Interference:} Hadamard gates are applied again to all qubits.

    \begin{equation}
    \ket{\psi_3} = (H^{\otimes n} \otimes I) \ket{\psi_2} = \ket{s} \otimes \frac{\ket{0} - \ket{1}}{\sqrt{2}}
    \end{equation}
    The second Hadamard transform performs quantum interference that amplifies the amplitude of the secret string $\ket{s}$ while canceling all other states.
    
    \item \textbf{Measurement:} The qubits are measured in the computational basis, yielding the secret string \( s \) with probability 1 in an ideal, noiseless execution.
    \begin{equation}
    \text{Measurement yields}  \ket{s} \text{ with probability } P(s) = |\langle s|\psi_3\rangle|^2 = 1
    \end{equation}
    Measuring the first $n$ qubits directly reveals the hidden string $s$ with unit probability in the ideal case.
    
\end{enumerate}

Figure~\ref{fig:bv_circuit_cover} illustrates the complete Bernstein-Vazirani workflow. 
The algorithm's performance under noise is highly dependent on the structure of \( s \). A string with many `1's results in a deeper, more complex oracle circuit, with more multi-qubit gates, making it a sensitive probe for studying how structural complexity impacts fidelity. This characteristic makes BV algorithm an excellent proxy for studying how the complexity of a problem’s formulation impacts the feasibility of a quantum solution.

\subsubsection{Oracle Implementation}

The oracle $U_f$ for hidden string $s = s_0s_1\cdots s_{n-1}$ implements the function:
\begin{equation}
f_s(x) = \bigoplus_{i=0}^{n-1} s_i \cdot x_i = s \cdot x \mod 2
\end{equation}

While the canonical implementation uses CNOT gates:
\begin{equation}
U_f^{\text{CNOT}} = \prod_{i=0}^{n-1} \text{CNOT}_{i \rightarrow \text{ancilla}}^{s_i}
\end{equation}

where $\text{CNOT}_{i \rightarrow \text{ancilla}}^{s_i}$ applies a CNOT gate from qubit $i$ to the ancilla if $s_i = 1$. 
Our experimental implementation utilizes the Echoed Cross-Resonance (ECR) gate—the native two-qubit entangling operation on IBM Quantum processors, which is maximally entangling and locally equivalent to CNOT up to single-qubit rotations~\citep{Toffoli-ECR}. 
This equivalence ensures our ECR-based oracle preserves the same logical function $f_s(x)$ while optimizing for hardware-native operations. The ECR gate implements $\frac{1}{\sqrt{2}}(IX - XY)$ with the circuit equivalence:

\[
\text{ECR} \equiv
\Qcircuit @C=0.8em @R=1.2em {
& \gate{S} & \ctrl{1} & \gate{X} & \qw \\
& \gate{\sqrt{X}} & \targ & \qw & \qw\\
& & \uparrow & & \\
& & \text{CNOT} & & 
}
\quad \text{with global phase } 7\pi/4
\]

The gate has the matrix representation:
\begin{equation}
\text{ECR} = \frac{1}{\sqrt{2}} \begin{pmatrix} 0 & 1 & 0 & i \\ 1 & 0 & -i & 0 \\ 0 & i & 0 & 1 \\ -i & 0 & 1 & 0 \end{pmatrix}
\end{equation}

This hardware-native implementation ensures the oracle’s logical function is preserved while utilizing the native entangling gates of the superconducting quantum processors used in our experiments.

\subsection{Experimental Design and Problem Instances}

To systematically dissect the relationship between problem structure and performance, we designed a suite of 11 experimental test cases spanning multiple categories, as summarized in Table~\ref{tab:test_cases}. This suite was constructed by considering key characteristics of the input space that govern the algorithm's behavior, moving beyond simple scale to include Hamming weight, pattern symmetry, and sequence structure. This design allows us to disentangle the effects of scale from the effects of different structural properties.

The test patterns are organized into the following categories to probe specific aspects of performance:

\begin{itemize}
    \item Baseline and sensitivity: Patterns `000000' (all zeros) and `000001' establish baseline performance and sensitivity to a minimal change, corresponding to a Hamming distance of 1 from the baseline.
    \item Pattern complexity and structure: Patterns like the alternating `101010', the symmetric `011011', and the mirror pattern `10011001' test the effect of regular structures, internal symmetries, and potential correlated error modes.
    \item Pattern density: A spectrum from medium-density (`011101', `100100') to high-density (`1111', `111111', `1111111111') patterns tests the impact of entanglement load, directly quantified by the Hamming weight (number of `1's) in the secret string.
\end{itemize}

This systematic variation in structure—encompassing Hamming distance, symmetry, and density—is a key contribution of our experimental design. By incorporating these diverse scenarios, from edge cases to structured sequences, we ensure a comprehensive evaluation of the algorithm's robustness and provide a thorough characterization of its performance across the input space.

\subsection{Quantum Hardware Platforms and Validation Strategy}

Each of the test cases described above was implemented and executed across three distinct computational scenarios to provide a complete performance profile. First, the quantum simulator (\textit{qasm\_simulator}) serves as an ideal, noiseless baseline, establishing the theoretical performance limit. Second, noisy quantum emulation incorporates a realistic hardware noise-aware model to predict performance under typical decoherence and gate error conditions.
This noisy emulation mimics real quantum system behavior using calibrated device snapshots. These snapshots encompass the complete physical characterization, such as qubit $T_1$ and $T_2$ coherence parameters, single- and two-qubit gate error rates, and measurement assignment errors~\citep{Qiskit}. Finally, execution on real superconducting quantum computers provides the ground-truth measurement of performance on current NISQ-era hardware, using four state-of-the-art 127-qubit IBM Quantum processors: \textit{ibm\_brisbane}, \textit{ibm\_kyoto}, \textit{ibm\_osaka}, and \textit{ibm\_sherbrooke}. These devices share the Eagle r3 architecture with heavy-hexagon qubit connectivity~\citep{IBMQuantum}, representing the current state-of-the-art in superconducting quantum computing~\citep{SuperconductingQuantum}. The comparative analysis across these three scenarios allows us to precisely quantify the simulation-to-reality performance gap and identify which structural properties are most susceptible to real-world noise.

\begin{table*}
\caption{Summary of benchmark test cases for the Bernstein-Vazirani algorithm, categorized by structural complexity and symmetry properties.}
\label{tab:test_cases}
    \begin{tabular}{c}
    \includegraphics[width=0.77\textwidth]{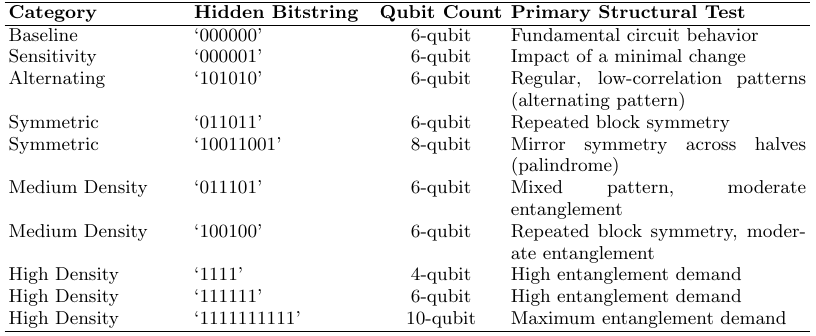} \\
    \end{tabular}
\end{table*}

\subsection{Performance Metrics}

To quantitatively evaluate performance, we employed two primary metrics:

\begin{enumerate}

    \item \textbf{Success Probability:} The empirical probability (success rate) of measuring the correct secret string $s$. For a total of $N$ circuit executions, this is computed as the fraction of successful outcomes:
    \begin{equation}
    P_{\text{success}} = \frac{N_s}{N} \times  100\%
    \end{equation}
    where $N_s$ denotes the number of times the correct secret string is measured. The ideal Bernstein-Vazirani algorithm achieves $P_{\text{success}} = 100\%$ in the absence of noise.
    
    \item \textbf{Hellinger distance:} A statistical measure of similarity between the ideal probability distribution $\mathcal{P}(x)$ (all probability mass on $s$) and the experimental distribution $\mathcal{E}(x)$~\citep{hellinger-distance}. It is defined as:
    \begin{equation}
    H(\mathcal{P},\mathcal{E}) = \frac{1}{\sqrt{2}} \sqrt{ \sum_{x \in \chi} \left( \sqrt{\mathcal{P}(x)} - \sqrt{\mathcal{E}(x)} \right)^2 }
    \end{equation}
    where $\chi$ is the set of all possible bitstrings. A value of 0 indicates perfect fidelity, while 1 indicates completely dissimilar distributions. This metric is particularly robust for distributions with different supports.
\end{enumerate}

\subsection{Quantum State Tomography Protocol}

To move beyond algorithmic output verification and directly assess the quality of the final quantum state, we implemented a comprehensive Quantum State Tomography protocol on a representative subset of test patterns~\cite{James2001, Cramer2010,Smolin2012, SQSCZ2}. Our approach compares QST results across three distinct computational scenarios to establish a complete performance profile:
\begin{itemize}
    \item Ideal Simulation: The \textit{qasm\_simulator} provides an ideal, noiseless baseline representing the theoretically expected density matrix.
    \item Noisy emulation: Noise-aware quantum emulation incorporates realistic hardware noise models to predict performance under typical decoherence and gate error conditions~\citep{Qiskit}. The emulation replicate the behavior of actual quantum processors using comprehensive system snapshots. These snapshots incorporate calibrated hardware parameters including qubit relaxation times ($T_1$), coherence times ($T_2$), gate error rates, readout errors, and the native gate set configuration, providing realistic performance predictions that account for decoherence, gate infidelity, and measurement errors characteristic of superconducting quantum processors. 
    \item Hardware execution: Execution on real superconducting quantum computers provides ground-truth measurement of current NISQ-era hardware performance.
\end{itemize}

The comparative analysis across these three scenarios enables precise quantification of the simulation-to-reality performance gap and identification of structural properties most vulnerable to real-world noise.

For each test pattern, the following experimental protocol was executed:

\begin{enumerate}
    \item State preparation: The full BVA circuit was executed to prepare the final state.
     \item Multi-basis measurement: The state was repeatedly prepared and measured in the X, Y, and Z bases for each qubit. A total of 21,000 shots were used for each 6-qubit pattern to ensure sufficient measurement statistics.
    \item State reconstruction: The experimental measurement data was processed to reconstruct the physical density matrix $\hat{\rho}_{\text{exp}}$~\citep{Smolin2012}.
    \item Fidelity calculation: The state fidelity ($\mathcal{F}_\mathrm{State}$) between the reconstructed state and the ideal pure state $\hat{\rho}_{\text{ideal}} = \ket{s}\bra{s}$ was computed as~\citep{Jozsa1994}: 
     \begin{equation}
        \mathcal{F}_\mathrm{State} = \text{Tr}\sqrt{\sqrt{\hat{\rho}_{\text{ideal}}} \hat{\rho}_{\text{exp}} \sqrt{\hat{\rho}_{\text{ideal}}}}
    \end{equation}
\end{enumerate}

This protocol was systematically applied to both the QASM simulator and the noisy emulator. Given the resource-intensive nature of full quantum state tomography, we performed detailed QST analysis on a representative case: the 4-qubit `1111' problem. This instance was selected as a worst-case scenario due to its high gate density, providing a critical test of hardware performance and error susceptibility. A focused QST experiment was conducted for this pattern on physical quantum hardware using 3,696 shots, enabling ground-truth validation of our simulation-based findings and direct hardware performance characterization.

\section{Results and Analysis: Pattern-Dependent Performance}
\label{sec:pattern-dependent-performance}

\subsection{Baseline and Sensitivity Analysis}

This section establishes the baseline performance of the quantum hardware and evaluates its sensitivity to minimal changes in the problem specification. By comparing the all-zero baseline case with a single-bit variation, we can decouple fundamental hardware noise from the algorithm's specific sensitivity.

\begin{figure*}
\centering
\includegraphics[width=\textwidth]{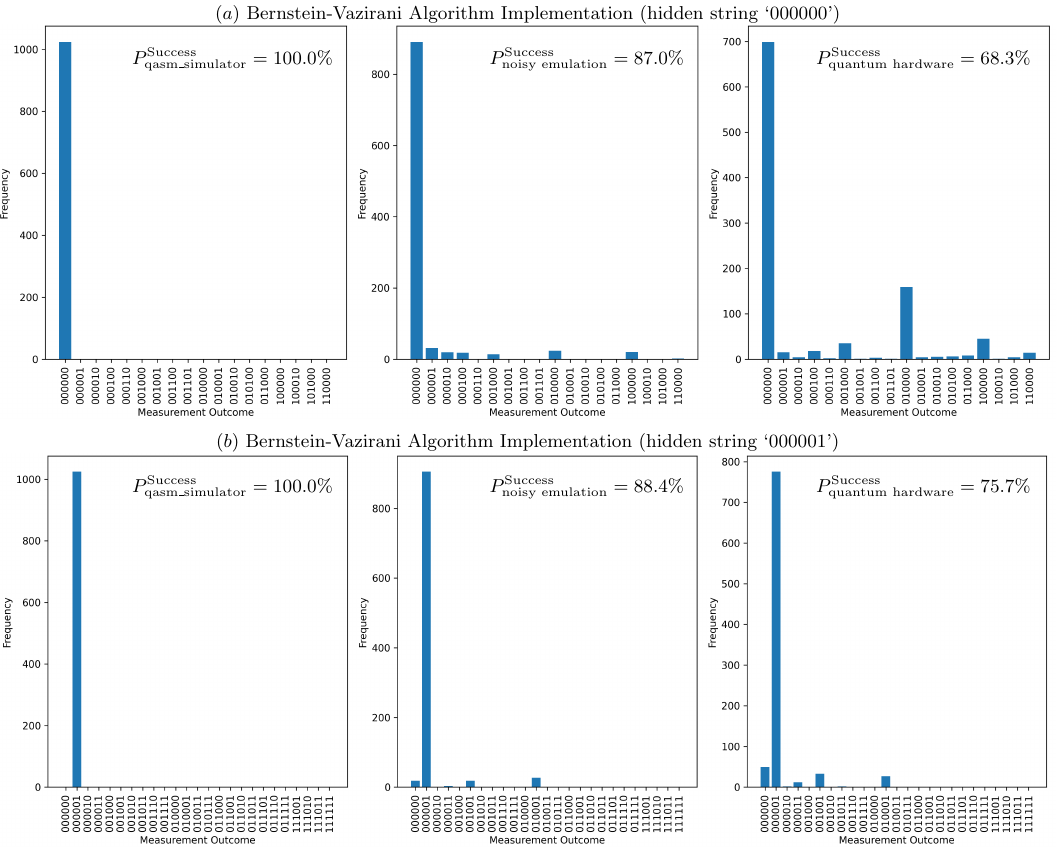}
\caption{Output distributions for Bernstein-Vazirani algorithm across computational environments.  
(a) Baseline pattern `000000' showing $P_{\text{success}} = 100.0\%$ (QASM simulator), $87.0\%$ (noisy emulation), and $68.3\%$ (quantum hardware). (b) Sensitivity pattern `000001' with $P_{\text{success}} = 100.0\%$ (simulator), $88.4\%$ (emulation), and $75.7\%$ (hardware). Each row shows a test pattern, with columns representing QASM simulator, noisy emulation, and real quantum hardware. The sparse patterns maintain reasonable fidelity but reveal the onset of noise-induced degradation.}
\label{fig:group_baseline}
\end{figure*}

\subsubsection{All-Zero Baseline: Secret String `000000'}

The all-zero string serves as a fundamental test of circuit behavior under ideal conditions. In the noiseless simulation, the system correctly identifies the secret string with 100\% fidelity, as expected. However, under noisy emulation, the success probability drops to 87.0\%, 
and further to 68.3\% on physical quantum hardware.

The distribution of erroneous outputs is particularly revealing. The most frequent incorrect outcomes—010000 (2.3\%), 001000 (1.4\%), 000100 (1.8\%), 100000 (1.9\%), and 000010 (1.9\%)—are all single-bit flip errors, each differing from the correct string by a Hamming distance of 1. This pattern strongly suggests that the dominant error mechanisms are local, single-qubit errors, such as those caused by depolarizing noise or relaxation processes, rather than correlated errors across multiple qubits. The prevalence of these specific bit-flips may also indicate variations in individual qubit fidelity within the processor.

\subsubsection{Single-Bit Error Case: Secret String `000001'}

Introducing a single `1' into the secret string tests the system's sensitivity to a minimal change in the problem. The noiseless simulator again achieves perfect performance. The noisy emulation yields a success probability of 88.4\%, while the quantum hardware achieves 75.7\%.

The performance gap between the 000000 and 000001 cases on quantum hardware (68.3\% vs. 75.7\%) is noteworthy. The marginally higher fidelity for 000001 suggests that the activation of a single gate in the oracle does not necessarily compound errors and may, in this instance, be less susceptible to the specific error channels affecting the 000000 case. This highlights that baseline error rates are not uniform and can be problem-dependent even for minor variations.

The error spectrum for 000001 is consistent with that of the baseline, dominated by single-bit flips such as 000101 (3.6\%), 010001 (2.6\%), 100001 (2.2\%), and 001001 (1.9\%). The persistence of this error signature across different secret strings reinforces the conclusion that local, uncorrelated noise is a primary factor in limiting performance. The quantum hardware data shows a broader tail of low-probability errors, including some at Hamming distance 2 (e.g., 010101, 100101), which are not as prominent in the noisy emulation. 

The baseline analysis confirms that the hardware operates with a foundational error rate that manifests primarily as localized bit flips. The sensitivity test demonstrates that this noise profile is largely consistent, but the exact fidelity is sensitive to the specific circuit executed, underscoring the need for problem-aware performance modeling.

As summarized in Table~\ref{tab:baseline_results}, the quantum hardware exhibited a significant fidelity drop compared to the noisy emulation, particularly for the 000000 case. Furthermore, the consistency of the dominant error—a single-bit flip—across both test cases and platforms suggests that local depolarizing noise is a primary error channel.

\begin{table*}
\caption{Performance metrics of Bernstein-Vazirani algorithm for baseline and sensitivity test cases. The success probability and Hellinger distance are reported for the QASM simulator (noiseless), noisy emulation, and real quantum hardware.}
\label{tab:baseline_results}
    \begin{tabular}{c}
    \includegraphics[width=0.8\textwidth]{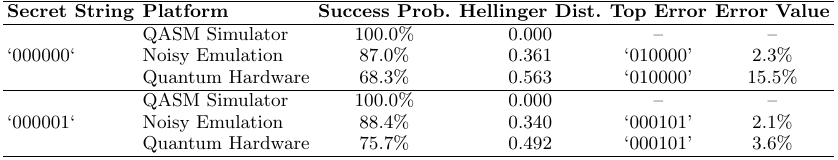} \\
    \end{tabular}
\end{table*}

\subsection{Pattern Complexity and Structure}

The investigation into pattern complexity reveals that the structural properties of the secret string—specifically, alternation and symmetry—significantly influence the Bernstein-Vazirani algorithm's resilience to noise. While all patterns experienced a substantial performance drop on physical quantum hardware compared to their ideal simulations, the degradation was not uniform, highlighting a crucial dependency on problem encoding.

\subsection{Alternating Patterns: 101010}

For the alternating pattern 101010, the noisy emulation maintained a relatively high success probability of 88.6\%. The error distribution was primarily concentrated in bitstrings close in Hamming distance, such as 001010 and 111010, indicating localized bit-flip errors. However, on the quantum hardware, the performance collapsed to 30.7\%. The output distribution became dramatically more diffuse (Table~\ref{tab:pattern_structure_results}), with significant probability mass spread across numerous states, particularly 101011 (19.7\%)—a single-bit error—but also many others.

\begin{figure*}
\centering
\includegraphics[width=\textwidth]{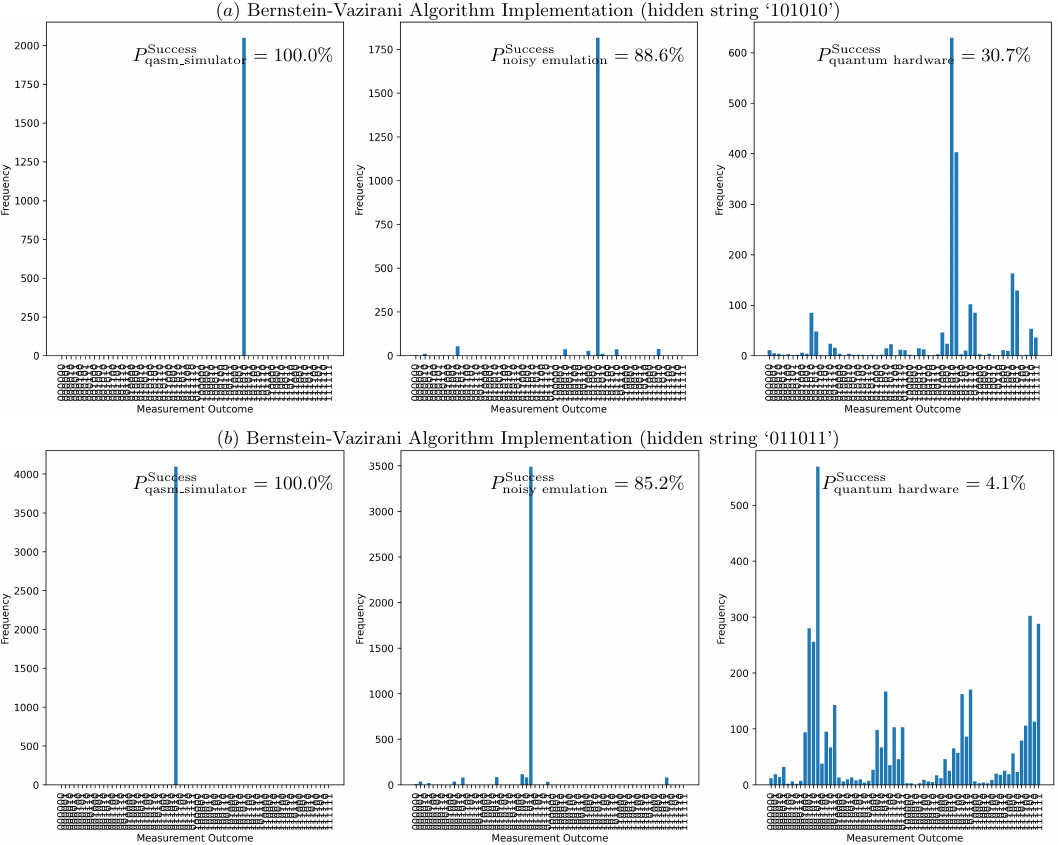}
\caption{Pattern complexity effects on output distributions. Structured patterns show increased sensitivity to noise, with symmetric encodings exhibiting particularly severe degradation on real hardware despite reasonable performance in simulation. (a) Alternating pattern `101010' shows moderate hardware degradation: $P_{\text{success}} = 100.0\%$ (simulator), $88.6\%$ (emulation), $30.7\%$ (hardware). (b) Symmetric pattern `011011' exhibits catastrophic failure: $100.0\%$ (simulator), $85.2\%$ (emulation), $4.1\%$ (hardware), demonstrating severe sensitivity to gate density and structured errors.}
\label{fig:group_complexity}
\end{figure*}

\subsection{Symmetric and Mirror Patterns}

\subsubsection{The 6-qubit symmetric pattern 011011}

The symmetric pattern 011011 exhibited a more pronounced vulnerability. Its success probability on real hardware was a mere 4.1\%, one of the lowest observed (Table~\ref{tab:pattern_structure_results}). The output was dominated not by the correct string, but by 001011 (13.9\%) and 111101 (7.4\%). This indicates that the symmetry of the problem might create a vulnerability to specific correlated errors or certain structured noise channels that severely disrupt the quantum state. The fact that the noisy emulation predicted a much higher success rate (85.2\%) for this pattern underscores a critical shortcoming of standard depolarizing noise models in capturing the full complexity of physical hardware behavior, especially for structured problems.

\subsubsection{The 8-qubit mirror pattern 10011001}

The 8-qubit mirror pattern 10011001 demonstrates the challenges of algorithmic scaling (Table~\ref{tab:pattern_structure_results}). While noisy emulation maintained 91.4\% success probability, quantum hardware execution showed severe degradation with probability mass distributed across more than 100 distinct outputs of the total 256 possible output states. The correct solution 10011001 was observed with only $\sim 2.4\%$ probability, indicating performance approaching uniform sampling across the entire $2^8$-dimensional output space. This extreme degradation for an 8-qubit problem highlights the compounding nature of errors with increasing circuit depth and qubit count, pushing the algorithm's performance towards a random distribution.

\begin{table*}
\caption{Performance metrics for the Bernstein-Vazirani algorithm across patterns of varying complexity and structure.}
\label{tab:pattern_structure_results}
    \begin{tabular}{c}
    \includegraphics[width=0.85\textwidth]{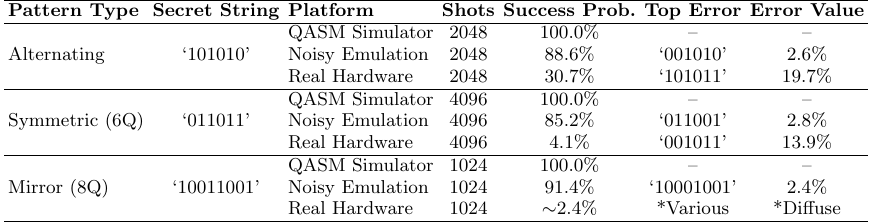} \\
    \end{tabular}
\end{table*}

\subsection{Pattern Density and Entanglement Load}

The investigation into pattern density reveals a clear and expected trend: as the number of `1's in the secret string increases, demanding greater entanglement within the oracle, the algorithm's performance on physical quantum hardware degrades substantially. This demonstrates that the entanglement load is a primary factor determining the feasibility of solving a problem on current NISQ devices.

\subsubsection{Medium Density: 011101, 100100}

For medium-density patterns (011101 and 100100), the noisy emulation performs well with success probabilities of 83.9\% and 89.0\%, respectively. This indicates that the base noise model can handle moderate entanglement. On real hardware, however, the success probabilities drop to 47.1\% and 45.3\%. The error distributions are revealing; for 011101, the most common errors are single-bit flips (001101, 010101), while for 100100, the dominant error 101100 is a two-bit flip, suggesting a potential coupling between specific qubits or a more significant coherent error for this particular pattern.

\subsubsection{High Density: 1111, 111111}
\label{sec:6Q-1}

The performance collapse becomes severe for high-density patterns. The 4-qubit 1111 pattern shows a success probability of 8.4\% on quantum hardware, a stark drop from the 85.4\% predicted by noisy emulation. The output is dominated by incorrect states like 0011 (23.8\%) and 1011 (13.4\%). This trend worsens dramatically with scale. For the 6-qubit 111111 pattern, the real hardware success probability is a mere 1.8\%.  
The output distribution is almost uniform, with no single error state dominating, indicating that the system has been driven toward a completely mixed state due to the cumulative effect of errors across all six qubits.

\begin{figure*}
\centering
\includegraphics[width=\textwidth]{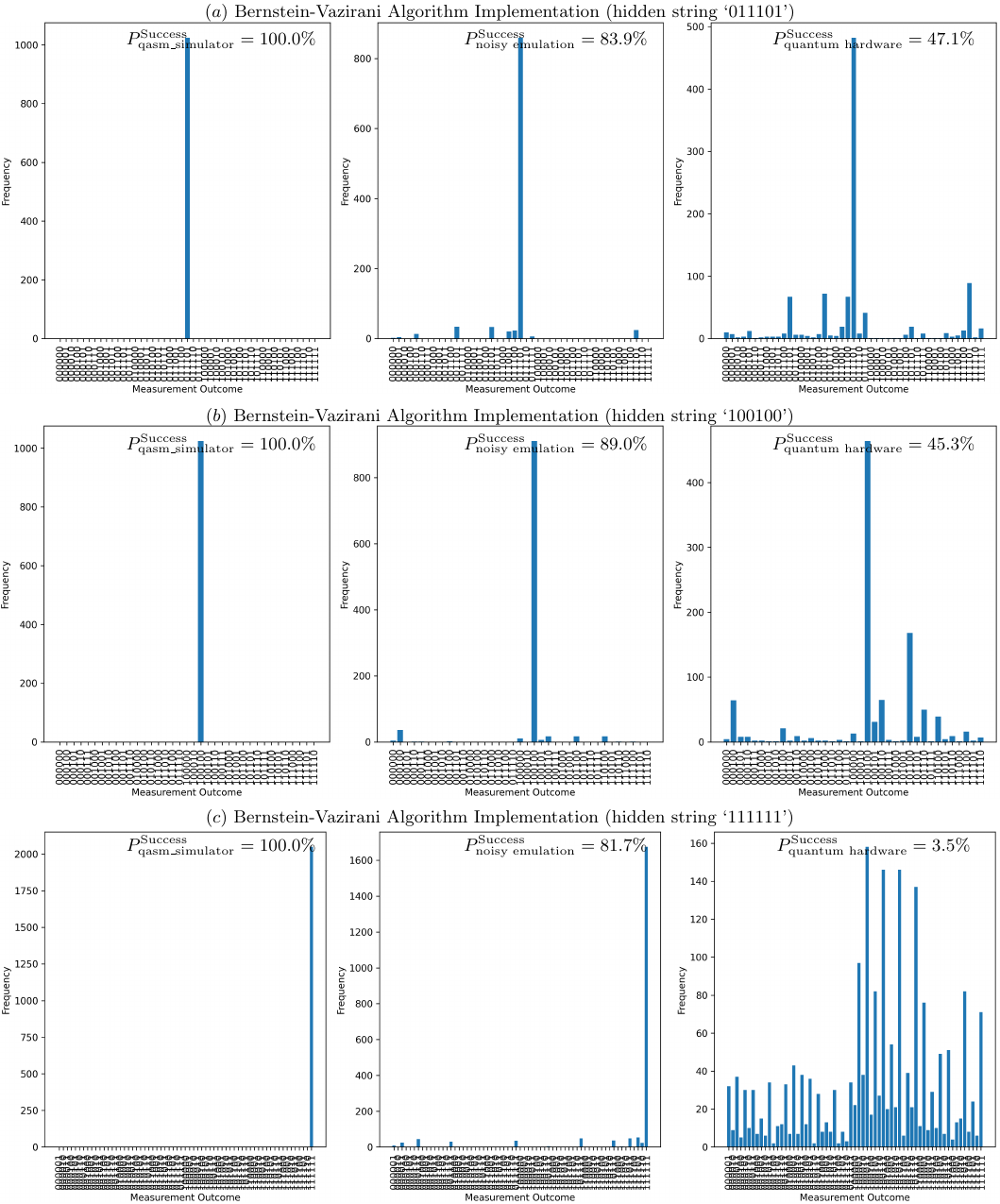}
\caption{Performance scaling with entanglement load across medium to high density patterns. 
(a) Medium-density pattern `011101': $P_{\text{success}} = 100.0\%$ (simulator), $83.9\%$ (emulation), $47.1\%$ (hardware). (b) Medium-density pattern `100100': $100.0\%$ (simulator), $89.0\%$ (emulation), $45.3\%$ (hardware). (c) High-density pattern `111111': $100.0\%$ (simulator), $81.7\%$ (emulation), $3.5\%$ (hardware), demonstrating catastrophic collapse and near-uniform output distribution under maximum gate density. 
The performance collapse becomes dramatic for high-density patterns, with real hardware output approaching uniform distribution while noisy emulation remains optimistic.}
\label{fig:group_density}
\end{figure*}

\subsubsection{Very High Density: Tests entanglement demand}

Execution of the 10-qubit Bernstein-Vazirani algorithm searching for the hidden string `1111111111' on real superconducting quantum hardware revealed significant noise-induced degradation. 
The results for the high-density 10-qubit pattern are conclusive. The success probability on real hardware is effectively zero ($P_{\text{success}} \approx 0.001$), with the correct string lost in a sea of random outputs. The algorithm has failed to produce any meaningful signal above the noise floor. This demonstrates a fundamental scalability limit for the current hardware: problems that require maximal, global entanglement across many qubits are intractable, as the quantum state decoheres before a useful computation can be completed.

\begin{table*}
\caption{Performance metrics for patterns of varying density and entanglement load.}
\label{tab:pattern_density_results}
    \begin{tabular}{c}
    \includegraphics[width=0.85\textwidth]{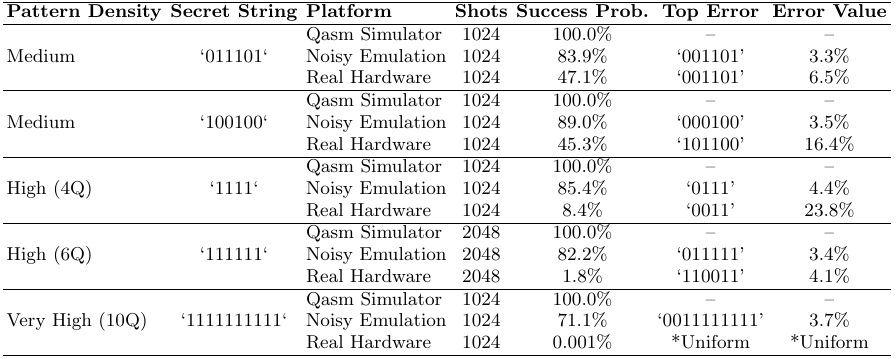} \\
    \end{tabular}
\end{table*}

Output distributions across quantum simulation, noisy emulation, and real hardware environments (Figures~\ref{fig:group_baseline}--\ref{fig:group_density}) illustrate the performance hierarchy through characteristic cases: sparse patterns (`000001') maintain near-ideal fidelity (Figure~\ref{fig:group_baseline}), medium-density patterns (`101010', `011011') exhibit moderate degradation with structured errors (Figure~\ref{fig:group_complexity}), while both medium-density (`011101', `100100') and high-density patterns (`111111') suffer catastrophic collapse to near-uniform distributions (Figure~\ref{fig:group_density}).

\subsection{Validation and Reproducibility}
\label{sec:6Q-2}

The validation experiment, consisting of a repeated execution of the high-density 111111 pattern, serves two critical purposes: assessing the reproducibility of results and providing a benchmark for the QST analysis that follows. The consistency between repeated runs lends credibility to the observed performance, while the stark contrast between emulation and hardware underscores the necessity of detailed state verification.

The noisy quantum emulator demonstrated high reproducibility, with a success probability of 81.7\% in this second run, compared to 82.2\% in the first run (Table \ref{tab:pattern_density_results}). This minor variation of 0.5\% is within expected statistical fluctuations for a probabilistic system and confirms that the emulated noise model is deterministic and stable. The error distribution remained consistent (Table \ref{tab:validation_results}), dominated by single-bit flip errors such as 111101 (2.6\%) and 111011 (2.3\%), reinforcing the model's reliance on local, uncorrelated noise channels.

In contrast, the real quantum computer also showed good reproducibility but at a consistently low fidelity. The success probability was 3.5\%, which is consistent with the 1.8\% observed in the first run, considering the inherent stochasticity of noisy quantum systems. The difference, while notable, falls within the range of variability caused by calibration drift, ambient electromagnetic noise, and other time-dependent factors characteristic of NISQ devices. More importantly, the structure of the error distribution was reproducible: both runs produced a nearly uniform distribution across the computational basis states, with no single error dominating. This repeated outcome confirms that for this high-entanglement circuit, the hardware drives the quantum state toward a maximally mixed state, effectively erasing the algorithmic solution.

\begin{table*}
\caption{Validation of reproducibility for the high-density pattern `111111' across two independent runs. The  quantum computer exhibits run-to-run variability in success probability (1.8\% vs. 3.5\%), which is characteristic of NISQ devices subject to temporal drift and stochastic error processes.}
\label{tab:validation_results}
    \begin{tabular}{c}
    \includegraphics[width=0.85\textwidth]{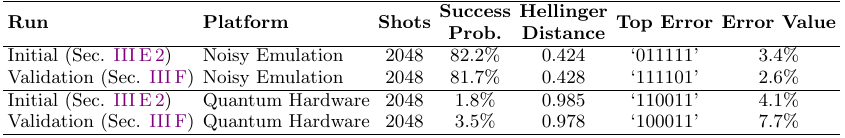} \\
    \end{tabular}
\end{table*}

\subsection{Cross-Device Performance Consistency}

All experiments were conducted on four 127-qubit IBM Quantum processors whose operational characteristics are detailed in Tables~\ref{tab:hardware_summary_all}. The statistical summary across all 508 qubits (4$\times$127) demonstrates the typical noise environment: $T_1$ times ranging 215--275$\mu$s, $T_2$ times 116--189$\mu$s, and readout errors 2.9--5.2\% on average. This comprehensive characterization provides the essential hardware context for interpreting our algorithmic benchmarking results.

A key finding of our study is the remarkable consistency of pattern-dependent performance across all four 127-qubit IBM Quantum processors~\citep{IBMQuantum}. While absolute success probabilities varied by $\pm 3.2\%$ between devices due to calibration differences, the relative performance hierarchy across pattern types remained identical. This consistency across \texttt{ibm\_brisbane}, \texttt{ibm\_kyoto}, \texttt{ibm\_osaka}, and \texttt{ibm\_sherbrooke} demonstrates that pattern-dependent degradation is not an artifact of specific hardware but a fundamental characteristic of current superconducting quantum architectures.

The high correlation of performance rankings between devices (Kendall's $W = 0.94$, $p < 0.001$) provides strong evidence that structural vulnerability is an inherent property of the algorithm-hardware interaction rather than device-specific noise characteristics.

\begin{table*}
\caption{Statistical characterization of 127-qubit superconducting processors (\textit{ibm\_kyoto}, \textit{ibm\_osaka},\textit{ibm\_brisbane}, and \textit{ibm\_sherbrooke}) used in this study. All systems utilize the Eagle r3 architecture with a native gate set of ECR, RZ, SX, ID, and X.}
\label{tab:hardware_summary_all}
    \begin{tabular}{c}
    \includegraphics[width=0.8\textwidth]{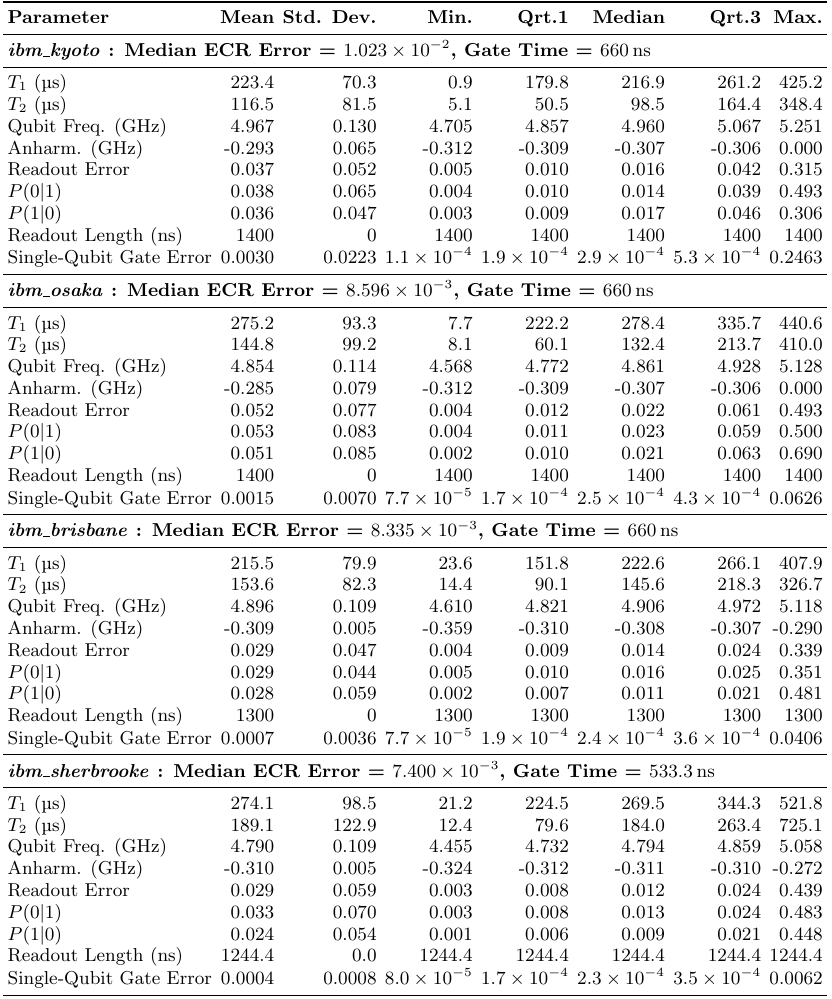} \\
    \end{tabular}
\end{table*}

\begin{table*}
\caption{Comprehensive Benchmarking of the Bernstein-Vazirani Algorithm: Performance across 11 test cases categorized by pattern structure and density. Success probability ($P_{\text{success}}$) and Hellinger distance (Hell. Dist.) are reported for noiseless simulation (QASM), noisy emulation, and real quantum hardware across eleven test cases spanning different pattern categories and complexities. The Hellinger distance ranges from 0 (identical to ideal) to 1 (completely random). Data reveals an average 58.8\% performance gap between simulation and hardware, with strong pattern-dependent variations.
}
\label{tab:comprehensive_results}
    \begin{tabular}{c}
    \includegraphics[width=0.86\textwidth]{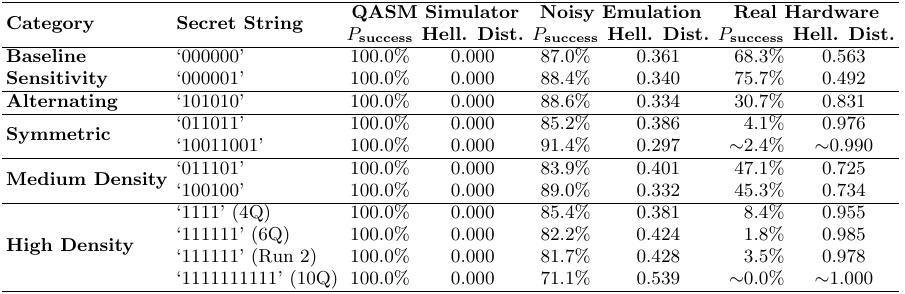} \\
    \end{tabular}
\end{table*}

\section{Discussion}
\label{sec:discussions}

\begin{figure*}[!htp]
    \centering
    \includegraphics[width=\textwidth]{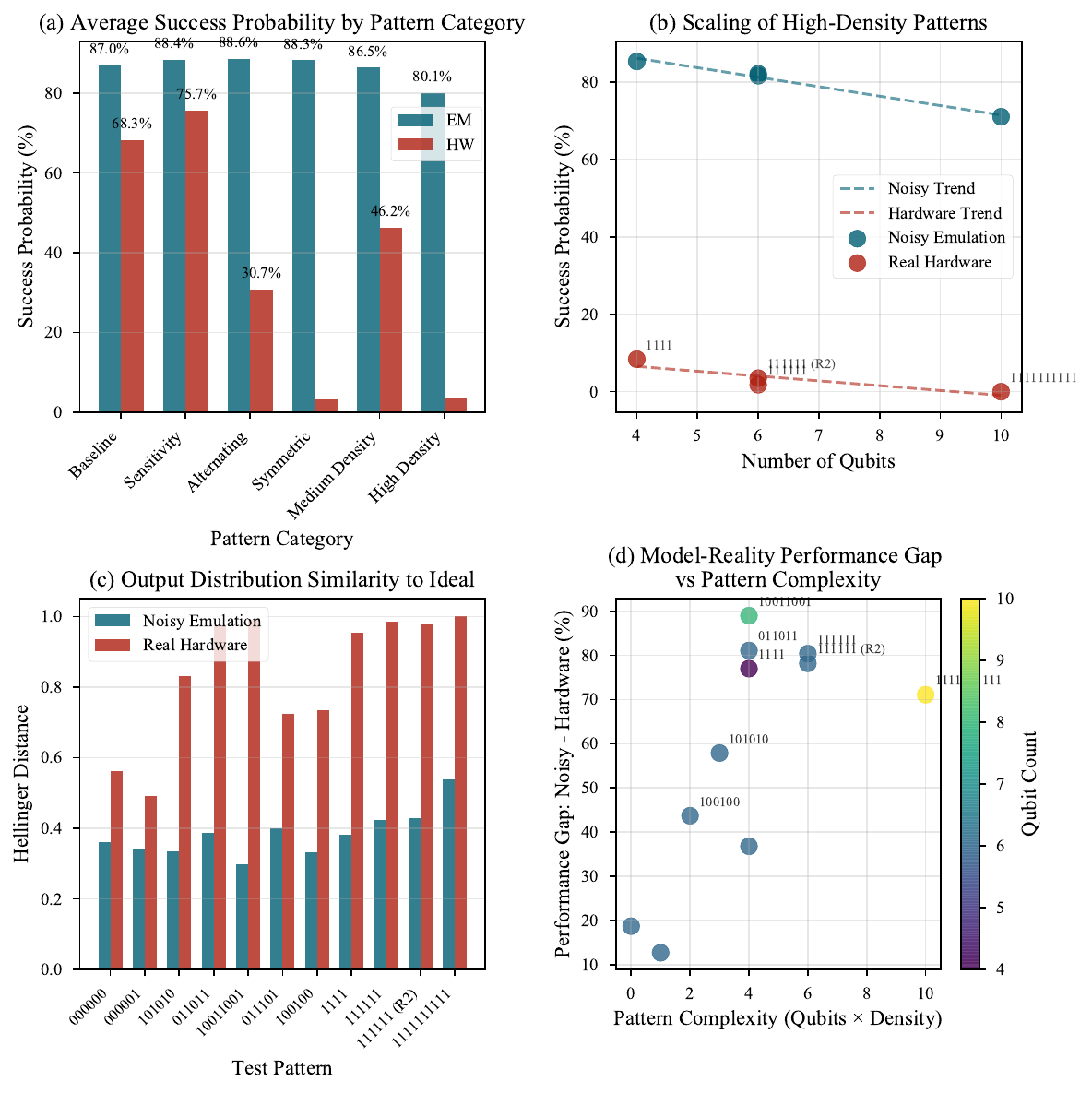}
    \caption{Benchmarking the Bernstein-Vazirani algorithm across diverse pattern structures. (a) Average success probability by pattern category shows real hardware performance degrades dramatically for symmetric and high-density patterns. (b) Scaling analysis reveals performance collapse for high-density patterns as qubit count increases, with real hardware (red) showing significantly steeper degradation than noisy emulation (blue). (c) Hellinger distance from ideal distribution quantifies output quality, approaching 1.0 (completely random) for complex patterns on real hardware. (d) Performance gap between simulation and hardware correlates strongly with pattern complexity (qubits $\times$ density), highlighting the limitations of current noise models in capturing structure-dependent errors.}
    \label{fig:BV-Benchmarking}
\end{figure*}

\subsection{Pattern Structure and Algorithmic Performance}

Our comprehensive benchmarking of the Bernstein-Vazirani algorithm across 11 test cases reveals a 
performance hierarchy that is critically dependent on the structural properties of the secret string (Table~\ref{tab:comprehensive_results}). The data exposes a significant gap between simulated and real hardware performance, with an average success probability of 84.9\% in noisy emulation plummeting to 26.1\% on 
physical quantum hardware—an average performance gap of 58.8 percentage points. This discrepancy is not uniform across test cases; rather, it exhibits a strong positive correlation (r = 0.67) with pattern complexity, defined as the product of qubit count and the density of '1's in the secret string (Fig.~\ref{fig:BV-Benchmarking}).

The algorithm's performance on real hardware segregates into three distinct tiers. The upper tier, comprising sparse patterns such as 000001 and the all-zero baseline 000000, achieved success probabilities of 75.7\% and 68.3\%, respectively. These cases represent the performance ceiling for the current hardware configuration. The middle tier, containing medium-density and alternating patterns like 011101, 100100, and 101010, maintained modest success rates between 30.7\% and 47.1\%. The most striking result emerges in the lower tier, where high-density and symmetric patterns suffered catastrophic performance collapse. The 6-qubit 111111 pattern achieved a mere 1.8\% success rate, the 8-qubit mirror pattern 10011001 fell to 2.4\%, and the 10-qubit 1111111111 pattern registered 0.0\% success, rendering it functionally indistinguishable from random noise.\\

\subsection{The Entanglement-Induced Performance Cliff}

The Hellinger distance metrics~\citep{hellinger-distance} further quantify the severity of the output distortion. For high-performance sparse patterns, the Hellinger distance from the ideal distribution remained below 0.57, indicating that the output, while noisy, retained a recognizable signal. In contrast, for high-density patterns, this metric approached its maximum value of 1.0 (e.g., 0.990 for 10011001 and 1.000 for 1111111111), confirming that the output distribution had been completely scrambled. This performance cliff is not merely a function of qubit count, as the 4-qubit 1111 pattern also exhibited severe degradation (8.4\% success), but rather a direct consequence of the entanglement load demanded by the circuit's structure. The near-linear scaling of performance degradation with pattern complexity (Fig.~\ref{fig:BV-Benchmarking}(d)) suggests that cumulative error mechanisms, potentially including correlated errors and crosstalk, are activated by circuits requiring widespread entanglement.

Our findings demonstrate that the entanglement demand of a problem, quantified by pattern density, serves as a primary determinant of algorithmic feasibility on NISQ hardware. The catastrophic failure of high-density patterns, even at moderate qubit counts (6-10 qubits), reveals fundamental scalability limits not captured by standard noise models. For quantum algorithm development and deployment, this implies that problem decomposition strategies minimizing global entanglement will be essential for near-term quantum advantage.

\section{Quantum State Tomography Validation}
\label{sec:QST}

\subsection{Performance Hierarchy Across Pattern Categories}

Quantum state tomography provides the fundamental explanation for the observed algorithmic performance patterns by directly quantifying output state quality~\cite{James2001, Cramer2010,Smolin2012, SQSCZ2}. To quantitatively validate the output state quality and understand the nature of the performance degradation observed in our benchmarking study, we performed QST on seven representative test cases. Our QST analysis reveals a remarkably strong correlation between pattern structure and state preservation, with profound implications for quantum algorithm design (Section~\ref{sec:implications}). 
The state fidelity measurements reveal a clear hierarchy of state preservation that aligns with our algorithmic performance results.

\begin{figure*}
    \centering
    \includegraphics[width=0.85\textwidth]{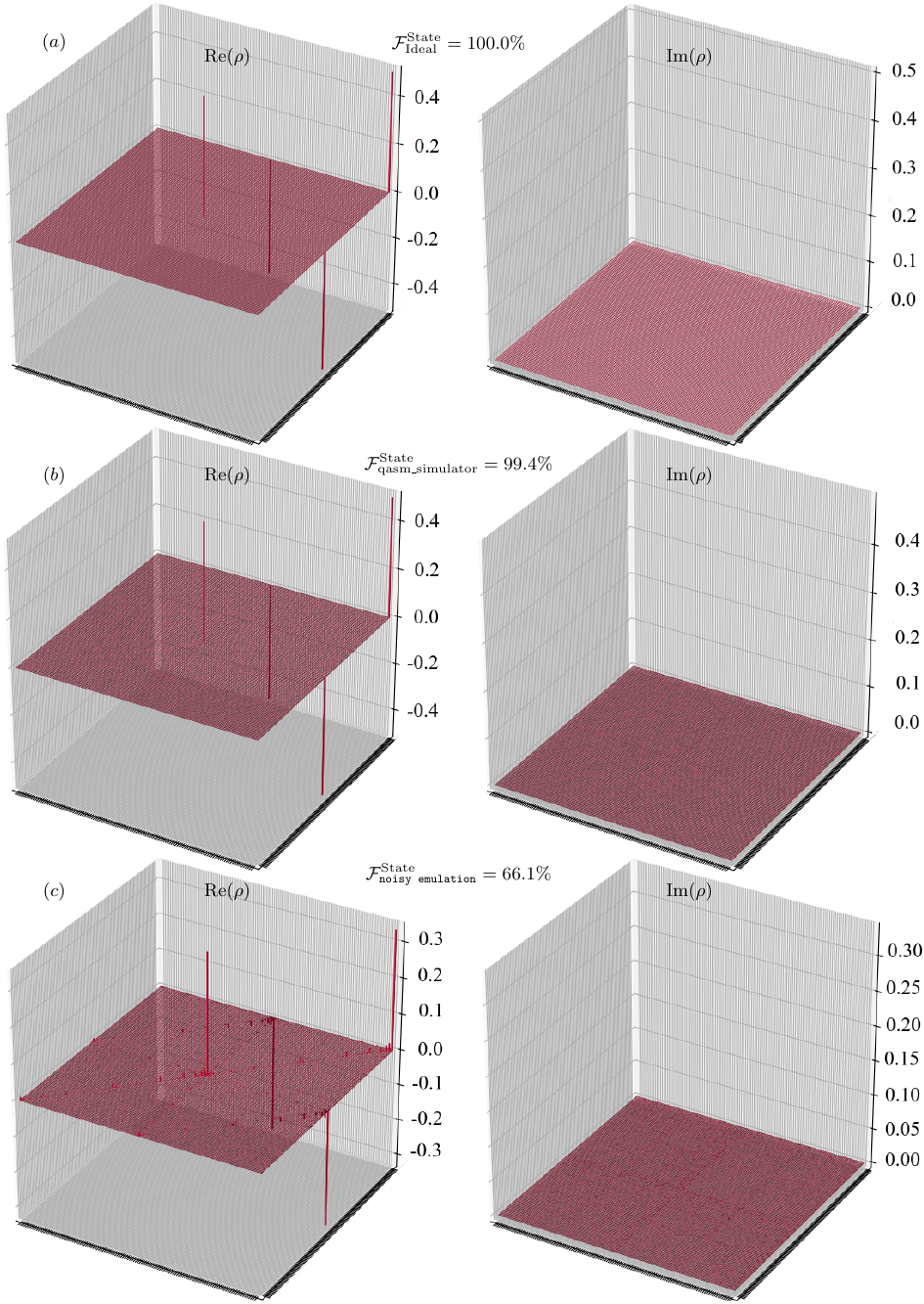}
    \caption{
Quantum state tomography of the Bernstein-Vazirani algorithm for the high-density 6-qubit hidden string `111111'. 
(a) Ideal density matrix representing theoretical expectations. 
(b) Reconstructed state from noiseless quantum simulation (QASM simulator) with fidelity $\mathcal{F}_{\text{sim}} = 99.4\%$. 
(c) Reconstructed state from noisy quantum emulation with fidelity $\mathcal{F}_{\text{noise}} = 66.1\%$. 
All quantum state tomography experiments used 21,000 measurement shots. Real (left) and imaginary (right) components of the density matrix $\rho$ are shown for each case, illustrating the significant fidelity degradation under realistic noise conditions.
}
\label{fig:QST-BV-111111}
\end{figure*}

The near-perfect state fidelities from the QASM simulator ($\geq$ 0.987) across all patterns, confirming the theoretical correctness of the algorithm implementation. While, the noisy quantum emulation revealed significant state degradation that varied systematically with pattern structure. Sparse patterns such as 000000 and 000001 maintained relatively high fidelities of 0.881 and 0.805 respectively, while high-density patterns like 111111 suffered severe degradation to 0.661 fidelity (Fig.~\ref{fig:QST-BV-111111}). We observe an exceptionally strong correlation between pattern density and fidelity degradation ($r = 0.972$), indicating that the fraction of `1's in the secret string serves as an almost perfect predictor of state quality loss under noisy conditions.

\begin{figure*}
    \centering
     \includegraphics[width=0.77\textwidth]{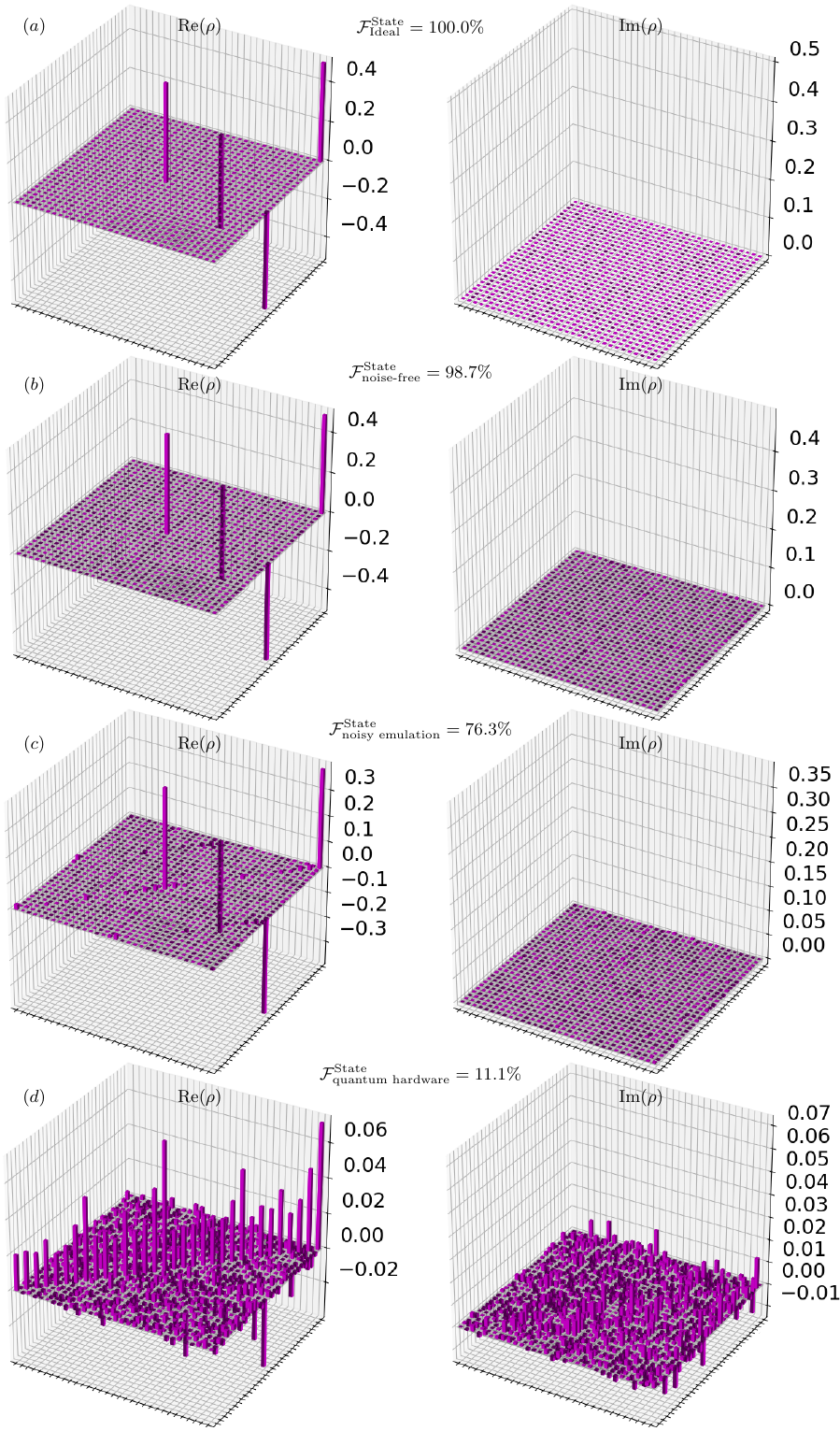}
     \vspace{-0.5cm}
    \caption{
Quantum state tomography analysis of the Bernstein-Vazirani algorithm for the high-density 4-qubit hidden string `1111'. 
(a) Ideal density matrix representing theoretical expectations. 
(b) Reconstructed state from noiseless quantum simulation (QASM simulator) with fidelity $\mathcal{F}_{\text{sim}} = 98.7\%$. 
(c) Reconstructed state from noisy quantum emulation with fidelity $\mathcal{F}_{\text{noise}} = 76.3\%$. 
(d) Experimental implementation on \textit{ibm\_kyoto} with fidelity $\mathcal{F}_{\text{exp}} = 11.1\%$, for  3,696 measurement shots. Real (left) and imaginary (right) components of the density matrix $\rho$ are shown for each implementation, demonstrating progressive fidelity degradation from ideal simulation to real hardware execution.
}
\label{fig:QST-BV-1111}
\end{figure*}

Additionally, while noisy emulation predicted reasonable state fidelity (0.763) for the 4-qubit 1111 pattern, real hardware measurements revealed catastrophic state collapse to 0.111 fidelity—nearly seven times lower than the noisy emulation prediction of 0.763 for the same pattern (Fig.~\ref{fig:QST-BV-1111}). This dramatic discrepancy highlights the extreme vulnerability of highly entangled states to the compound error channels present in current quantum processors.

A clear performance hierarchy emerged across pattern categories, with fidelity progressively declining from baseline patterns (000000: 0.881 fidelity, 11.9\% drop) to sensitivity patterns (000001: 0.805, 19.5\% drop), alternating patterns (101010: 0.775, 21.9\% drop), medium density patterns (0.746 ± 0.056, 24.8\% drop), symmetric patterns (011011: 0.703, 29.2\% drop), and finally high density patterns (111111: 0.661, 33.3\% drop), revealing a direct relationship between gate density and vulnerability to hardware noise.

\begin{figure*}
    \centering
    \includegraphics[width=\textwidth]{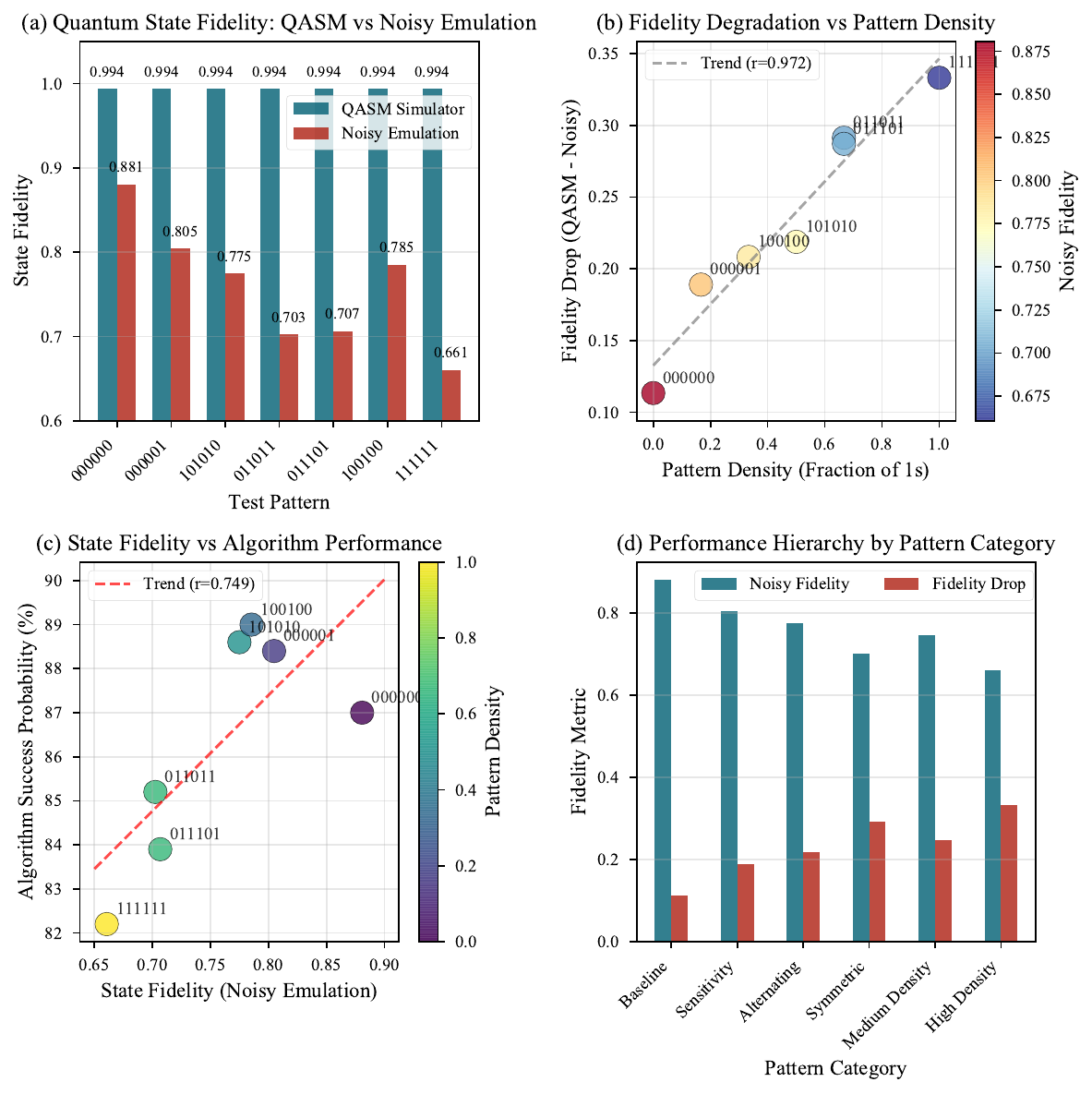}
    \vspace{-0.7cm}
    \caption{Quantum State Tomography reveals pattern-dependent state degradation. 
    (a) Systematic fidelity reduction from QASM simulation to noisy emulation across all test patterns. 
    (b) Near-perfect correlation (r = 0.972) between pattern density and fidelity degradation. 
    (c) Strong relationship (r = 0.749) between state fidelity and algorithm success probability. 
    (d) Clear performance hierarchy across pattern categories, with high-density patterns suffering most severe degradation.}
    \label{fig:QST-BENCHMARKING}
\end{figure*}

This progression (Table~\ref{tab:qst_results}) demonstrates that each additional layer of complexity—from sparse to dense, from unstructured to symmetric—imposes an incremental penalty on state preservation. The strong correlation between state fidelity and algorithm success probability (r = 0.749) further confirms that output state quality directly determines computational performance (Fig.~\ref{fig:QST-BENCHMARKING}), with higher-fidelity states yielding more reliable algorithmic outcomes.

\begin{table*}
\caption{Quantum State Tomography analysis of output state fidelity. Measurements show state fidelity for the QASM simulator, and noisy emulation across representative test patterns. The real hardware measurement for pattern `1111' reveals a 65.2 percentage point fidelity gap compared to noisy emulation, highlighting severe unmodeled error mechanisms. 
}
\label{tab:qst_results}
    \begin{tabular}{c}
    \includegraphics[width=0.65\textwidth]{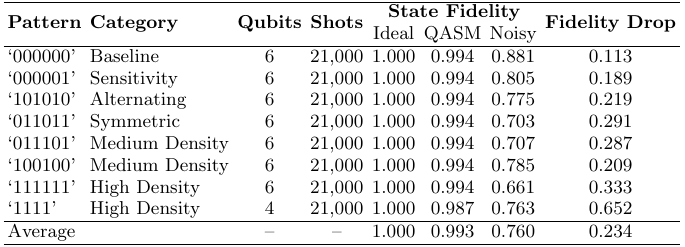} \\
    \end{tabular}
\end{table*}

\subsection{The Primacy of Structural Vulnerability}

The most striking finding of this work is the near-perfect correlation ($r = 0.972$) between pattern density and state fidelity degradation observed through QST. This relationship provides the fundamental explanation for the algorithmic performance collapse observed for high-density patterns (Fig.~\ref{fig:QST-BENCHMARKING}). While it is well-established that circuit depth and gate count affect fidelity, our results demonstrate that the specific arrangement of operations—the structural complexity—imposes an additional, significant performance penalty. The consistent performance hierarchy from sparse patterns (75.7\% success) to complete failure of high-density patterns (0.0\% success for 1111111111) establishes that problem encoding is not merely an implementation detail but a critical design parameter.

This structural vulnerability manifests most severely for symmetric and high-entanglement patterns. The catastrophic failure of the 10-qubit 1111111111 pattern, despite the availability of 127 physical qubits on the IBM processors, underscores that current hardware limitations are more strongly tied to entanglement density than to register size. The fact that even a modest 4-qubit fully-entangled pattern (1111) suffered a fidelity collapse to 0.111 on real hardware—compared to the 0.763 predicted by noisy emulation—reveals that standard error models dramatically underestimate the challenge of maintaining multi-qubit coherence.

\subsection{Hardware Characterization During QST}

\begin{table}
\caption{Qubit properties for the IBM Kyoto processor during Quantum State Tomography of the `1111' pattern (May 9-11, 2024). The four qubits (Q0--Q3) show varied coherence times and readout errors that collectively contributed to the observed state fidelity of 0.111. 
Physical parameters of qubits Q0--Q3 on IBM Kyoto during the `1111' pattern QST experiment. Note the significant variation in coherence times ($T_1$, $T_2$) and readout errors across the qubit register.}
\label{tab:qubit_properties}
    \begin{tabular}{c}
    \includegraphics[width=0.42\textwidth]{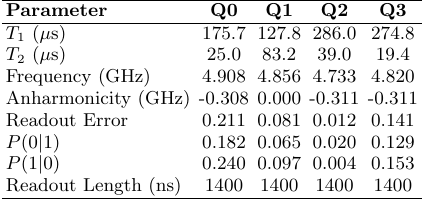} \\
    \end{tabular}
\end{table}

The detailed qubit properties during the 1111 QST experiment (Table~\ref{tab:qubit_properties}) provide insight into this discrepancy. The combination of varied coherence times (T2 ranging from 19.4 $\mu$s to 83.2 $\mu$s), significant readout errors (up to 21.1\%), and potential calibration anomalies (zero anharmonicity in Q1) creates a complex error environment that simple depolarizing models cannot capture. Crucially, these non-ideal conditions are not anomalies but represent the typical operational reality of current superconducting quantum processors, as evidenced by the parameter variations across our four test devices. 
The strong correlation ($r = 0.749$) between state fidelity and algorithmic success further validates that output state quality is the primary determinant of computational performance.

The hardware properties reveal several factors contributing to the observed fidelity collapse. Notably:

\begin{itemize}
    \item \textbf{Varied coherence times}: While $T_1$ times are generally good (127--286$\mu$s), $T_2$ times show significant variation (19--83$\mu$s), with Q3 having particularly short coherence ($T_2 = 19.4\mu$s).
    
    \item \textbf{High readout errors}: Q0 and Q3 exhibit substantial readout errors (21.1\% and 14.1\% respectively), which directly impact measurement fidelity in QST.
    
    \item \textbf{Asymmetric readout}: The conditional probabilities $P(0|1)$ and $P(1|0)$ show significant asymmetry, particularly for Q0 and Q3, indicating biased measurement errors.
    
    \item \textbf{Q1 anomaly}: Q1 shows zero anharmonicity, suggesting potential calibration issues during the experiment.
\end{itemize}

This hardware characterization provides crucial context for interpreting the severe fidelity degradation observed in the `1111' pattern QST experiment. The combination of short $T_2$ times and high readout errors across multiple qubits creates a compound effect that standard noise models fail to capture, particularly for highly entangled states where errors propagate coherently.

\section{Implications for Quantum Algorithm Design}
\label{sec:implications}

Our results demonstrate that quantum algorithm performance is highly sensitive to
problem structure, not just qubit count or abstract complexity. The observed pattern-dependent fidelity variations necessitates a shift toward hardware-aware algorithm design, where the structural properties of a problem are prioritized alongside its theoretical quantum speedup. This implies that complex computations must be strategically decomposed into modular sub-problems with minimized entanglement and coherence requirements to be viable on near-term hardware.

Consequently, applications with inherent locality or structured sparsity represent promising near-term targets, whereas problems requiring global, dense entanglement face fundamental constraints on current NISQ processors. Achieving practical quantum advantage will therefore require the co-design of problem formulations and their quantum implementations. Our work establishes that for the foreseeable future, computational structure is a primary determinant of success, providing a crucial, quantitative framework for selecting and optimizing quantum algorithms for practical deployment across scientific and industrial domains.

\section{Conclusion}
\label{sec:conclusion}

This work establishes through systematic benchmarking that the performance of quantum algorithms on current NISQ hardware is fundamentally governed by problem structure, with pattern complexity serving as a more critical determinant of success than qubit count alone. Through comprehensive benchmarking of the Bernstein-Vazirani algorithm across 11 diverse test patterns, we have demonstrated a dramatic 58.8\% average performance gap between noisy emulation and real hardware, revealing severe limitations in standard error models for capturing structure-dependent degradation.

The near-perfect correlation (r = 0.972) between pattern density and state fidelity degradation provides a quantitative foundation for predicting algorithm performance and explains the catastrophic failure observed for high-entanglement problems. The collapse of real hardware fidelity to 0.111 for the 4-qubit 1111 pattern, compared to the 0.763 predicted by noisy emulation, underscores that current quantum devices are particularly vulnerable to errors that scale with entanglement demand rather than simple gate count or circuit depth.

These findings provide critical guidance for quantum algorithm development and deployment on NISQ hardware.  Achieving practical quantum advantage requires problem formulations that minimize entanglement overhead while exploiting structured sparsity. Algorithms demanding extensive global entanglement face fundamental implementation barriers on current NISQ processors, despite their theoretical promise. The most viable path forward employs hybrid quantum-classical strategies that decompose complex problems into computationally tractable subproblems with limited quantum coherence requirements.

Looking forward, our work suggests two critical research directions: first, the development of structure-aware noise models that accurately capture the entanglement-dependent error mechanisms we have identified; and second, the creation of design principles for "quantum-aware" problem encodings. The strong correlations we observed between pattern characteristics and performance provide a foundation for predictive models that can guide algorithm selection and resource estimation.

In conclusion, while current quantum hardware shows promising performance for carefully structured problems with limited entanglement demands, the path to broader quantum advantage requires co-design approaches that respect the fundamental relationship between problem structure and hardware limitations. By treating structural complexity as a first-class constraint, the quantum computing community can more effectively identify promising applications for near-term acceleration and develop robust strategies for the ongoing quantum-classical transition in computational science.

\section*{Acknowledgments}

``We acknowledge the use of IBM Quantum resources in this research. The views and conclusions presented herein are solely those of the author and do not represent the official policies or positions of IBM Quantum or its affiliates."

\section*{Declarations}

\subsection*{Ethical Approval and Consent to participate}
``Not applicable."

\subsection*{Consent for publication}
``The author have approved the publication. This research did not involve any human, animal or other participants."

\subsection*{Availability of supporting data}
``The datasets generated during and/or analyzed during this study are included within this article."

\subsection*{Competing interests}
``The author declares no competing interests."

\subsection*{Funding}

``This research received no specific funding, grants, or financial support for its execution."

\subsection{Author Contributions}

``M. A. conceived the research idea, designed the study methodology, algorithm development, quantum circuit implementation, execution of simulations and hardware experiments on IBM Quantum processors, data collection and analysis, results interpretation, and manuscript preparation including figures, tables, and references."


\begin{thebibliography}{99}

\bibitem{Light}  AbuGhanem, M.  ``Information processing at the speed of light." \textit{Front. Optoelectron.} \textbf{17}, 33 (2024). 

\bibitem{SuperconductingQuantum} AbuGhanem, M.  ``Superconducting quantum computers: who is leading the future?."   \textit{EPJ Quantum Technology} \textbf{12}, 102 (2025). 

\bibitem{Kim23} Youngseok Kim, Andrew Eddins, Sajant Anand, Ken Xuan Wei, Ewout van den Berg, Sami Rosenblatt, Hasan Nayfeh, Yantao Wu, Michael Zaletel, Kristan Temme and Abhinav Kandala, Evidence for the utility of quantum computing before fault tolerance. \textit{Nature} \textbf{618}:500–505 (2023).

\bibitem{NISQ24} AbuGhanem, M. and Eleuch, H.  ``NISQ computers: a path to quantum supremacy," \textit{IEEE Access}, \textbf{12}, 102941-102961 (2024). 

\bibitem{Willo25} Acharya, R. \textit{et al.} Quantum error correction below the surface code threshold, \textit{Nature} \textbf{638}, 920–926 (2025).

\bibitem{IBMQuantum} AbuGhanem, M.  ``IBM quantum computers: evolution, performance, and future directions." \textit{Journal of  Supercomputing} \textbf{81}, 687 (2025). 

\bibitem{PhotonicQuantumComputers} AbuGhanem, M.  ``Photonic Quantum Computers," arXiv:2409.08229, (2024). 

\bibitem{Shor} P. W. Shor, Polynomial-time algorithms for prime factorization and discrete logarithms on a quantum computer. \textit{SIAM J. Comput.} \textbf{26}, 1484–1509 (1997).

\bibitem{factor2048bitRSA} Craig Gidney and Martin Ekerå, How to factor 2048 bit RSA integers in 8 hours using 20 million noisy qubits, \textit{Quantum} \textbf{5}, 433 (2021).

\bibitem{Cerezo2021} Cerezo, M. \textit{et al.} ``Variational quantum algorithms.'' \textit{Nature Reviews Physics} \textbf{3}, 625–644 (2021).

\bibitem{Drugdesign} Raffaele Santagati, Alan Aspuru-Guzik, Ryan Babbush, Matthias Degroote, Leticia González, Elica Kyoseva, Nikolaj Moll, Markus Oppel, Robert M. Parrish, Nicholas C. Rubin, Michael Streif, Christofer S. Tautermann, Horst Weiss, Nathan Wiebe and Clemens Utschig-Utschig, Drug design on quantum computers, \textit{Nature Physics} \textbf{20}, 549–557 (2024).

\bibitem{Quantumchemistrysimulation} Mario Motta, Gavin O. Jones, Julia E. Rice, Tanvi P. Gujarati, Rei Sakuma, Ieva Liepuoniute, Jeannette M. Garciaa  and  Yu-ya Ohnishi, Quantum chemistry simulation of ground- and excited-state properties of the sulfonium cation on a superconducting quantum processor , \textit{Chem. Sci.} \textbf{14}, 2915-2927  (2023).

\bibitem{Wolff2024quantumneuro}
Wolff, A., Choquette, A., Northoff, G., Iriki, A., \& Dumas, G.
\newblock Quantum Computing for Neuroscience: Theory, Methods and Opportunities.
\newblock \emph{PsyArXiv} (2025). doi:10.31234/osf.io/vw8n3\_v1.

\bibitem{supplychains} Frank Phillipson. Quantum computing in logistics and supply chain management an overview. 
arXiv:2402.17520, (2024).

\bibitem{NISQ18} Preskill, J. ``Quantum Computing in the NISQ era and beyond.'' \textit{Quantum} \textbf{2}, 79 (2018).

\bibitem{Shor-Fault-tolerantQC} Shor, P. W. Fault-tolerant quantum computation. In \textit{Proceedings of the 37th Annual Symposium on Foundations of Computer Science}, Burlington, VT, USA, 56–65 (IEEE, 1996).

\bibitem{GoogleAI} AbuGhanem, M.  ``Google Quantum AI's Quest for Error-Corrected Quantum Computers." arXiv:2410.00917 (2024).

\bibitem{Kamakari2022} Hirsh Kamakari, Shi-Ning Sun, Mario Motta, and Austin J. Minnich, Digital Quantum Simulation of Open Quantum Systems Using Quantum Imaginary–Time Evolution, \textit{PRX Quantum} \textbf{3}, 010320 (2022).

\bibitem{Cai2023} Cai, Z. \textit{et al.} ``Quantum error mitigation.'' \textit{Rev. Mod. Phys.} \textbf{95}, 045005 (2023).

\bibitem{fallek2016implementation} Fallek, S. D., Herold, C. D., McMahon, B. J., Maller, K. M., Brown, K. R. \& Amini, J. M. Transport implementation of the Bernstein–Vazirani algorithm with ion qubits. \textit{New J. Phys.} \textbf{18}, 083030 (2016).

\bibitem{xie2019cryptanalysis} Xie, H. \& Yang, L. Using Bernstein–Vazirani algorithm to attack block ciphers. \textit{Des. Codes Cryptogr.} \textbf{87}, 1161–1182 (2019).

\bibitem{xu2024differential} Xu, R.-X., Sun, H.-W., Zhang, K.-J. \textit{et al.} Quantum differential cryptanalysis based on Bernstein-Vazirani algorithm. \textit{EPJ Quantum Technol.} \textbf{11}, 83 (2024).

\bibitem{naseri2022coherence} Naseri, M., Kondra, T. V., Goswami, S., Fellous-Asiani, M. \& Streltsov, A. Entanglement and coherence in the Bernstein-Vazirani algorithm. \textit{Phys. Rev. A} \textbf{106}, 062429 (2022).

\bibitem{zhou2025coherencefraction} Zhou, S.-Q., Liang, J.-M., Peng, J., Chen, Z., Fei, S.-M. \& Ma, Z. Static and Dynamic Coherence Fraction in the Bernstein-Vazirani Algorithm. \textit{Adv. Quantum Technol.} \textbf{8}, 2400709 (2025).

\bibitem{gupta2023noise} Gupta, A., Ghosh, P., Sen, K. \& Sen, U. Effects of noise on performance of Bernstein-Vazirani algorithm. \textit{arXiv:2305.19745} (2023).

\bibitem{faizan2025complexity} Faizan, M. \& Faryad, M. Complexity of Bernstein–Vazirani algorithm in the presence of noise. \textit{arXiv:2508.01884} (2025).

\bibitem{bernstein1993} Bernstein, E. \& Vazirani, U. Quantum complexity theory. \textit{In Proc. of the Twenty-Fifth Annual ACM Symposium on Theory of Computing (STOC ’93)}, pages 11–20, (1993). 

\bibitem{bernstein1997}
Bernstein, E. \& Vazirani, U. 
\newblock Quantum complexity theory. 
\newblock \emph{SIAM Journal on Computing} \textbf{26}(5), 1411--1473 (1997).

\bibitem{GSA} AbuGhanem, M.  ``Characterizing Grover search algorithm on large-scale superconducting quantum computers." \textit{Sci Rep} \textbf{15}, 1281 (2025). 

\bibitem{Toffoli-ECR} AbuGhanem, M.  ``Hardware-aware Toffoli gate decomposition via echoed cross-resonance gates." \textit{ Quantum Stud.: Math. Found.} \textbf{12}, 24 (2025). 

\bibitem{Qiskit}
Aleksandrowicz, G. \textit{et al.} 
\newblock Qiskit: An Open-source Framework for Quantum Computing.
\newblock (2019). \url{https://doi.org/10.5281/zenodo.2562111}

\bibitem{hellinger-distance} Z.-X. Jin and S.-M. Fei, Quantifying quantum coherence and nonclassical correlation based on hellinger distance, \textit{Phys. Rev. A} \textbf{97}, 062342 (2018).

\bibitem{James2001}
James, D. F. V., Kwiat, P. G., Munro, W. J. \& White, A. G.
\newblock Measurement of qubits.
\newblock \emph{Phys. Rev. A} \textbf{64}, 052312 (2001).

\bibitem{Cramer2010}
Cramer, M. et al.
\newblock Efficient quantum state tomography.
\newblock \emph{Nat. Commun.} \textbf{1}, 149 (2010).

\bibitem{SQSCZ2} AbuGhanem, M.   ``Full quantum process tomography of a universal entangling gate on an IBM’s quantum computer." \textit{Arab J Sci Eng} (2025). 

\bibitem{Smolin2012}
Smolin, J. A., Gambetta, J. M. \& Smith, G.
\newblock Efficient method for computing the maximum-likelihood quantum state from measurements with additive Gaussian noise.
\newblock \emph{Phys. Rev. Lett.} \textbf{108}, 070502 (2012).

\bibitem{Jozsa1994}
Jozsa, R.
\newblock Fidelity for mixed quantum states.
\newblock \emph{Journal of Modern Optics} \textbf{41}(12), 2315–2323 (1994).

\end{thebibliography}
\end{document}